\documentclass[11pt,a4paper]{article}
\pdfoutput=1
\usepackage{jcappub}

\usepackage{color}
\input{colordvi.tex}

\usepackage{amsmath}
\usepackage{graphicx}
\usepackage{float}
\usepackage{amsmath}

\usepackage{amsmath,amssymb}
\usepackage{graphicx}
\usepackage{subfigure}
\usepackage{verbatim}
\usepackage{makeidx}
\usepackage{amssymb}
\usepackage{ulem}

\title{Decaying dark matter and the tension in $\sigma_8$}

\author[a]{Kari Enqvist,}
\author[a]{Seshadri Nadathur,}
\author[a]{Toyokazu Sekiguchi}
\author[b]{and Tomo Takahashi}
\affiliation[a]{University of Helsinki and Helsinki Institute of Physics, 
P.O. Box 64, FI-00014, Helsinki, Finland}
\affiliation[d]{Department of Physics, Saga University, Saga 840-8502, Japan}

\emailAdd{kari.enqvist@helsinki.fi}
\emailAdd{toyokazu.sekiguchi@helsinki.fi}
\emailAdd{seshadri.nadathur@helsinki.fi}
\emailAdd{tomot@cc.saga-u.ac.jp}

\abstract{
We consider decaying dark matter (DDM) as a resolution to the possible tension between cosmic microwave 
background (CMB) and weak lensing (WL) based  determinations of the amplitude of matter fluctuations, $\sigma_8$.
We perform N-body simulations in a model where dark matter decays into dark radiation and develop an accurate fitting formula for the
non-linear matter power spectrum, which enables us to test the DDM model by the combined measurements of CMB, WL and the baryon acoustic oscillation (BAO).
We employ a Markov chain Monte Carlo analysis to examine the overlap of posterior distributions of the cosmological parameters, comparing 
 CMB alone with WL+BAO. We find an overlap that is significantly larger in the DDM model than in the standard CDM model. 
This may be hinting at DDM, although current data is not constraining enough to unambiguously favour
a non-zero dark matter decay rate $\Gamma$. From the combined CMB+WL data, we obtain a lower bound 
$\Gamma^{-1}\ge 97$ Gyr at 95~\% C.L, which is less tight than the constraint from CMB alone.
}


\begin{document}

\begin{flushright}
\end{flushright}

\maketitle
\flushbottom

\section{Introduction} 
\label{sec:introduction}

Observations of temperature and polarization anisotropies in the 
cosmic microwave background (CMB) 
have been a milestone in cosmology, which allow precision tests of various cosmological models. 
The six-parameter $\Lambda$ Cold Dark Matter ($\Lambda$CDM) model has emerged as the concordance model of the Universe due to its ability to fit the CMB anisotropies and various other cosmological observations.\footnote{
See Ref.~\cite{Planck:2015xua} and references therein.
} The recent observation of CMB by Planck
allows us to determine the cosmological parameters in this model to sub-percent accuracy~\cite{Planck:2015xua}.

At the same time, closer analysis of the level of consistency of the CMB and other 
cosmological data has revealed a number of possible tensions in the $\Lambda$CDM model.
One that has recently received attention is the value of the amplitude of matter fluctuations, 
characterized by the parameter $\sigma_8$. 
The value of $\sigma_8$ estimated from observations of large-scale structure such as the weak lensing 
of distant galaxies and abundance of galaxy clusters
is significantly smaller than the one inferred from the Planck 
CMB power spectrum~\cite{Ade:2013zuv,Hamann:2013iba,Battye:2013xqa,Petri:2015ura}.\footnote{
Most of these studies were based on the Planck 2013 results, but the discrepancy still exists in the recent Planck 2015 data~\cite{Planck:2015xua}.
} This tension might be indicating a need to move beyond the standard $\Lambda$CDM and to search for a mechanism that would suppress the matter fluctuations at late times.

Several authors have argued that the tension can be alleviated if 
active neutrinos are massive or sterile neutrinos exist~\cite{Hamann:2013iba,
Battye:2013xqa,Beutler:2014yhv,MacCrann:2014wfa,Battye:2014qga}.\footnote{
The suppression of $\sigma_8$ has also been discussed recently in 
a specific model that combines dark matter and dark radiation~\cite{Buen-Abad:2015ova}.
} However, this is not the only possible solution for the tension, and other models should also be tested on equal footing.
A better understanding of effects of astrophysical processes might also be important (see, e.g., Ref.~\cite{Osato:2015lja}). 
 In this paper we explore another possibility, where dark matter can decay with a lifetime 
$\Gamma^{-1}$ of the order of the age of the Universe.

Such a decaying dark matter (DDM) model is 
viable if the decay products interact little with the visible particles; otherwise 
the lifetime should be much longer.\footnote{For a recent constraint, we refer to Ref.~\cite{Ando:2015qda}. }
In this paper, we restrict ourselves to the simplest class of DDM models, 
where decay products are invisible and massless. In addition, we also assume that all the dark matter
consists of DDM.
Such a model can be characterised with only one extra parameter $\Gamma$ in addition to
the standard six parameters of the $\Lambda$CDM model.
While one can consider more complicated models of DDM with, e.g., 
finite masses of decay products or a mixture of DDM and CDM,
our analysis of the simplest case 
is sufficient to demonstrate the potential of DDM models.

In DDM models, the formation of cosmic structure is in general suppressed.
This is because the decay products free-stream  with finite kick-velocity
and can escape from the gravitational potential wells surrounding matter over-densities.
On the other hand, at early times before significant decay has occurred, DDM models are completely degenerate
with the CDM case. This allows the amplitude of matter fluctuations 
at late time to be suppressed relative to the CDM case, without significantly affecting the 
CMB anisotropy spectrum except through the integrated Sachs-Wolfe effect.

To distinguish DDM from CDM using cosmological observations requires a precise
knowledge of the matter power spectrum 
over a wide range of scales. On large scales, structure formation in DDM models
can be described by linear perturbation 
theory and has been studied by many 
authors~\cite{Flores:1986jn,Takahashi:2003iu,Ichiki:2004vi,
Wang:2010ma,Aoyama:2011ba,Aoyama:2014tga,Audren:2014bca}.
Based on this linear approach and using the Planck CMB power spectrum combined with
measurements of baryon acoustic oscillations (BAO) and the galaxy power spectrum on linear scales, Ref.~\cite{Audren:2014bca} gives a lower bound $\Gamma^{-1}\ge 160$ Gyr
on the lifetime of DDM.\footnote{
Ref.~\cite{Aubourg:2014yra} gives a comparable constraint, $\Gamma^{-1}\ge 100$ Gyr,
using only information on distances.
}

On the other hand, to make use of other cosmological data requires capturing the behaviour of the 
matter power spectrum to sufficient accuracy on non-linear scales, which can be achieved 
through using N-body simulations.\footnote{
For semi-analytic studies based on the halo model see, e.g.,
Refs~\cite{Cen:2000xv,Oguri:2003nn,Wang:2010ma,Peter:2010au}.
} 
In this paper we modify the {\tt Gadget2} code~\cite{Springel:2000yr,Springel:2005mi}
to incorporate the effects of DDM. 

N-body simulations of DDM models
have recently been used to study the effects on halo inner profiles~\cite{Peter:2010jy,Cheng:2015dga}, 
the Lyman-$\alpha$ forest~\cite{Wang:2013rha} and 
effects on non-linear matter power spectrum~\cite{Cheng:2015dga}. 
The implications for small-scale problems in the CDM models have also been discussed in 
Ref.~\cite{Wang:2014ina}.
In these studies, decay products were assumed to be non-relativistic.
On the other hand, we focus on models with massless decay product, 
and follows the methodology developed in Ref.~\cite{Suto}.
We will show that our linear 
calculations and N-body simulations agree well
on scales where non-linearity is subdominant. 
We develop a fitting formula which can describe non-linear matter power spectrum well, 
using which we are able to obtain parameter constraints in the DDM model from observational data,
especially including the recent CFHTLens weak lensing measurement~\cite{Heymans:2013fya}. 

We note that the possibility for DDM to resolve the tension in the $\sigma_8$ estimations 
has also been discussed in Refs.~\cite{Aoyama:2014tga,Cheng:2015dga,Berezhiani:2015yta}, 
although the models and parameter regions of DDM they consider differ from ours.
Also, in this paper we focus on the weak lensing measurement, which has not been considered in previous DDM studies.

The structure of this paper is as follows. In Section~\ref{sec:linear}, we define
the DDM model and discuss the cosmological linear perturbation theory.
In Section~\ref{sec:Nbody}, we describe our N-body simulations and discuss the
effects of DDM on the non-linear matter power spectrum. Using the simulation results,
we obtain constraints on the DDM 
model from current cosmological observations in Section~\ref{sec:constraints}. 
In particular, we examine to what extent the DDM model
can mitigate the tension in the $\sigma_8$ estimations.
We conclude in the Section~\ref{sec:conclusion}. In Appendix~\ref{app:approx}, 
we describe the approximations used in the Boltzmann hierarchy
in the linear calculation. Appendix~\ref{app:fit} contains details of our fitting formula for the non-linear matter power spectrum.

Throughout this paper, we denote a derivative with respect to the conformal time $\tau$ 
by a dot, i.e. $\dot{}\equiv \partial/\partial \tau$. 
We will assume a flat Universe and adiabatic initial perturbations
with a power-law primordial power spectrum.

\section{Model and linear perturbation calculation} 
\label{sec:linear}

\subsection{Model of decaying dark matter}
\label{sec:model}
We consider a model of dark matter (DM) decaying into dark radiation (DR) with decay rate $\Gamma$.
We assume the DM is  non-relativistic. Note that as long as the decay products are massless, the number 
of DR particles generated from decay of single DM is irrelevant to the evolution of the cosmological perturbations.

Roughly speaking, current observations constrain the lifetime $\Gamma^{-1}$ to be larger than $100$~Gyr~\cite{Audren:2014bca,Aubourg:2014yra}, so we focus on values of $\Gamma^{-1}$ around 100~Gyr in obtaining observational constraints.

\subsection{Background evolution}
\label{sec:background}
Given a model of DDM, energy conservation for the DM and DR fluids
in the unperturbed background metric yields
\begin{eqnarray}
\dot{\bar\rho}_{dm}+3\mathcal H\bar\rho_{dm}&=&-a\Gamma\bar\rho_{dm},
\label{eq:bg1} \\
\dot{\bar\rho}_{dr}+4\mathcal H\bar\rho_{dr}&=&a\Gamma\bar\rho_{dm},
\label{eq:bg2}
\end{eqnarray}
where $\bar\rho_i$ is the unperturbed energy density of fluid $i$ and $\mathcal H\equiv \dot a/a$ is the conformal Hubble parameter.

Due to the energy transfer between  DM and DR fluids, $\bar\rho_{dm}$ and $\bar\rho_{dr}$ cannot be represented 
as explicit functions of $a$. This means that the present value of the Hubble parameter, $H_0=$100$h$ km/sec/Mpc, 
cannot be given as an input parameter but is rather a derived parameter obtained by solving the
differential equations~\eqref{eq:bg1} and \eqref{eq:bg2} 
in conjunction with the Friedmann equation 
\begin{equation}
\mathcal H^2=8\pi G a^2\sum_i \bar\rho_i/3~, 
\end{equation}
where $G$ is Newton's constant.

To specify the background evolution, it is convenient to introduce the following quantity:
\begin{equation}
\tilde \omega_i(a)=\frac{\bar\rho_i(a)a^{3(1+w_i)}}{\rho_{\rm crit}/h^2}, 
\end{equation}
where  $\rho_{\rm crit}/h^2\equiv3(100{\rm km/sec/Mpc})^2/8\pi G$ and  $w_i$ is the (constant) equation of state parameter of fluid $i$.
Given the initial values $\tilde \omega_i(a=0)\equiv\omega_i$, we can solve the differential equations \eqref{eq:bg1} and \eqref{eq:bg2} 
for $\tilde \omega_{dm}$ and $\tilde \omega_{dr}$ to obtain $\Omega_i h^2\equiv \tilde \omega_i(a=1)$ and $h$.
Note that if fluid $i$ does not interact with other fluids (i.e., if the  r.h.s. in Eqs. \eqref{eq:bg1} and \eqref{eq:bg2} vanish),
$\tilde \omega_i(a)$  is constant in time and $\omega_i=\Omega_i h^2$. 

The background cosmology is specified by the parameters
\begin{equation}
(\omega_b,~\omega_{dm},~\omega_\Lambda,~\Gamma), 
\end{equation}
where the subscripts $b$ and $\Lambda$ indicate respectively baryons and dark energy.
We assume that initially $\omega_{dr}=0$ so that DR is produced only through the decay of DM.
Given these parameters, $\Omega_{dm} h^2$, $\Omega_{dr} h^2$ and $h=\sqrt{\sum_i\Omega_i h^2}$ can be obtained as derived parameters.
For later convenience, we define the reduced Hubble constant in the limit of no decay $\Gamma=0$, 
to be $h_\emptyset=\sqrt{\sum_i\omega_i}$.
In this section and in Section~\ref{sec:Nbody}, we adopt a fiducial set of initial density parameters:
\begin{equation}
(\omega_b=0.02216,~\omega_{dm}=0.119,~\omega_\Lambda=0.318),
\label{eq:fiducial}
\end{equation}
and assume an amplitude $A_s=2.43\times10^{-9}$
and spectral index $n_s=0.963$ at a pivot scale $k=0.05$~Mpc$^{-1}$ for the primordial power spectrum.
This parameter set gives $h_\emptyset=0.677$ and
corresponds to the best-fit $\Lambda$CDM model from the Planck CMB temperature power spectrum combined with measurements of small-scale CMB power spectrum and BAO~\cite{Ade:2013zuv}.

In closing this section, in order to avoid confusion, let us clarify the difference between $\omega_{dm}$ and $\Omega_{dm}h^2$. 
As we noted, the density parameter $\tilde \omega_{dm}(t)$ is not constant and evolves with the time due to the decay of DM.
We have defined the initial and final values of $\tilde \omega_{dm}$ as 
$\omega_{dm}\equiv\tilde\omega_{dm}(a=0)$
and $\Omega_{dm}h^2\equiv\tilde\omega_{\rm}(a=1)$. In the limit of no decay $\Gamma=0$, $\omega_{dm}$ and $\Omega_{dm}h^2$
coincide. Let us also remind the readers of the definitions of $h_\emptyset\equiv\sqrt{\sum_i\omega_i}$ and 
$h\equiv\sqrt{\sum_i\Omega_ih^2}$. The former is the reduced hubble parameter that are realized in the limit of no decay, while the latter
is the actual reduced hubble parameter.

\subsection{Linear perturbation evolution and 
matter power spectrum}
\label{sec:perturbation}
Let us now discuss cosmological perturbation at linear order.
We follow the notation of Ref.~\cite{Ma:1995ey} and
treat the perturbed quantities in the synchronous gauge of DM.
In this gauge, the scalar part of the perturbed metric is 
\begin{equation}
ds^{2}=a(\tau)^{2}\left[d\tau^{2}+\left\{\delta_{ij}+h_{ij}(\vec x, \tau)\right\}dx^{i}dx^{j}\right],
\end{equation}
where $h_{ij}$ is given in Fourier space as 
\begin{equation} 
h_{ij}(\vec k, \tau)=
 h_{\rm L}(\vec k,\tau) 
\hat k_{i} \hat k_{j}
+6\eta_{\rm T} (\vec k,\tau) 
(\hat k_{i} \hat k_{j} - \dfrac{1}{3}\delta_{ij}), 
\end{equation}
and $\hat{}$ denotes a unit vector, i.e. $\hat k=\vec k/k$. 
In the following, we focus on a single Fourier mode and 
dependences on $\vec k$ will be abbreviated.

In this gauge, the continuity equation for DM is 
\begin{equation}
\dot\delta_{dm}=-\frac{\dot h_{\rm L}}2, 
\label{eq:cont_phi}
\end{equation}
where $\delta_{dm}=\delta \rho_{dm}/\bar \rho_{dm}$ is the fractional overdensity of DM.
Note that Eq.~\eqref{eq:cont_phi} is the same as in CDM.

On the other hand, since DR is massless, its perturbation equations can be given 
in terms of its brightness function defined as
\begin{equation}
X(\hat n, \tau)\equiv \frac1{4a^4\bar\rho_{dr}}\int \frac{q^2 dq}{2\pi^2}q\,\delta f_{dr}(\vec q,\tau),
\end{equation}
where $\delta f_{dr}(\vec q,a)$ is the perturbation in the phase-space distribution of DR
and $\hat n$ is the unit vector of a comoving momentum $\vec q=q\hat n$. 
The Boltzmann equation for DR can 
be given in terms of $X(\hat n, \tau)$ as \cite{Ichiki:2004vi}
\begin{equation}
\dot X+ik\mu X=
\left\{\dot\eta_T-\frac12(\dot h_L+6\dot \eta_T) \mu^2\right\}
-\frac{\dot{\tilde\omega}_{dr}}{\tilde\omega_{dr}}\left(X-\frac{\delta_{dm}}4\right), 
\label{eq:brightness0}
\end{equation}
where $\mu=\hat k\cdot\hat n$. 
After multipole expansion, Eq.~\eqref{eq:brightness0} is recast into
\begin{equation}
\dot X_l=\frac{k}{2l+1}
\left\{lX_{l-1}-(l+1)X_{l+1}\right\}
-\frac{\dot h_L}6\delta_{l0}
+\frac{\dot h_L+6\dot\eta_T}{15}\delta_{l2}
-\frac{\dot{\tilde\omega}_{dr}}{\tilde\omega_{dr}}
\left(X_l-
\frac{\delta_{dm}}4\delta_{l0}\right),
\label{eq:brightness}
\end{equation}
where $X(\hat n,\tau)=\sum^\infty_{l=0}(-i)^l(2l+1)X_l(\tau)P_l(\mu)$, with $P_l$ being the Legendre polynomial.

To solve the perturbation equations, we have modified the Boltzmann code {\tt CAMB}~\cite{Lewis:1999bs}.
We truncated the Boltzmann hierarchy at $\ell_{\rm max}=50$ up to moderately sub-horizon scales, $k\tau\le \ell_{\rm max}$.
However, there is a difficulty in solving the Boltzmann hierarchy Eq.~\eqref{eq:brightness} on deep sub-horizon scales.
A na\"ive truncation of the hierarchy yields fictitious reflections at $\l_{\rm max}$, and the accumulation of them
quickly blows up $X_l$. To avoid this, we adopted the approximations developed in Ref.~\cite{Blas:2011rf}, 
details of which are provided in Appendix~\ref{app:approx}.

\begin{figure}
  \begin{center}
      \hspace{0mm}\scalebox{1.2}{\includegraphics{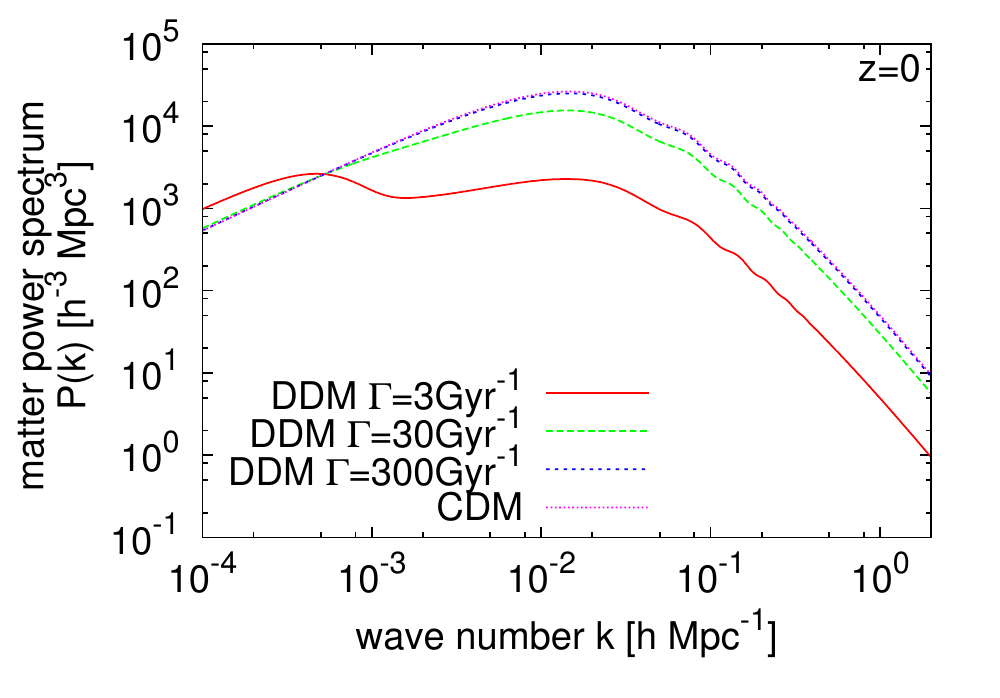}} \\
  \end{center}
  \caption{Power spectrum of non-relativistic matter today $(z=0)$, for 
DDM models  with $\Gamma$[Gyr$^{-1}$]=3 (red), 30 (green) and 300 (blue).
  For reference, the case with the standard CDM is also plotted (magenta).
  }  
  \label{fig:mpk}
\end{figure}
In Fig.~\ref{fig:mpk}, we plot the matter power spectrum $P(k)$ at present.
It is clearly seen that as $\Gamma$ is increased, $P(k)$ is suppressed
on small scales. We have also checked that  our code reproduces the CMB power spectrum shown in Ref.~\cite{Ichiki:2004vi}
and the matter power spectrum in Ref.~\cite{Audren:2014bca}. 
By adjusting the accuracy settings in {\tt CAMB}, we have determined that within the parameter range of interest the numerical accuracy of the linear calculation code is sub-percent
for the CMB temperature and polarization power spectra, and $\sim1~\%$ for the matter power spectrum.

\section{N-body simulation and non-linear matter power spectrum} 
\label{sec:Nbody}

\subsection{Simulation} 
\label{sec:simulation}

We perform N-body simulations of collisionless particles 
by using the public code {\tt Gadget2}
\cite{Springel:2000yr,Springel:2005mi}.
To incorporate effects of DDM, 
we make two modifications, which are also implemented previously in Ref.~\cite{Suto}.
One is that the mass of N-body particles $m$ is made time-dependent and varies as
\begin{equation}
m(t)=m_i\{(1-r_{dm})+r_{dm} e^{-\Gamma t}\},
\end{equation} 
where $m_i$ is the initial particle mass and $r_{dm}\equiv \omega_{dm}/(\omega_b+\omega_{dm})$
is the initial fraction of the non-relativistic matter in DDM.
The other is that the expansion of the Universe is given
by solving Eqs.~\eqref{eq:bg1} and \eqref{eq:bg2}.

In this treatment, effects of perturbations in DR are neglected. 
However, since DR is relativistic and free-streams, 
on sub-horizon scales --- where
Newtonian gravity and thus the N-body simulation are valid --- 
DR does not cluster and is highly homogeneous and isotropic.
On the other hand, on super-horizon scales, the perturbations
in DR cannot be neglected. However, on these scales, perturbations are 
linear and their evolution can be accurately computed within the perturbation theory treatment. 
Thus, by interpolating results of N-body simulations on small scales and
linear perturbation theory on large scales, we can predict the matter power spectrum
in the DDM model over a broad range of scales.

The initial matter power spectrum at redshift $z=49$ is computed using {\tt CAMB}, 
from which initial distributions of N-body particles are
generated using the {\tt 2LPTic} code~\cite{Crocce:2006ve}, which uses second-order Lagrangian perturbation theory.
The simulations contain 1024$^3$ particles, and are performed with three different box sizes, $L=1250$, 500 and 200~$h_\emptyset^{-1}$Mpc. For each box size, we generate three different realizations, and 
for each realization of the initial conditions, we run simulations with different values of $\Gamma$, which allows 
us to extract effects of the dark matter decay with statistical fluctuations minimised.

\subsection{Non-linear matter power spectrum} 
\label{sec:power}

\begin{figure}
  \begin{center}
      \hspace{0mm}\scalebox{1.3}{\includegraphics{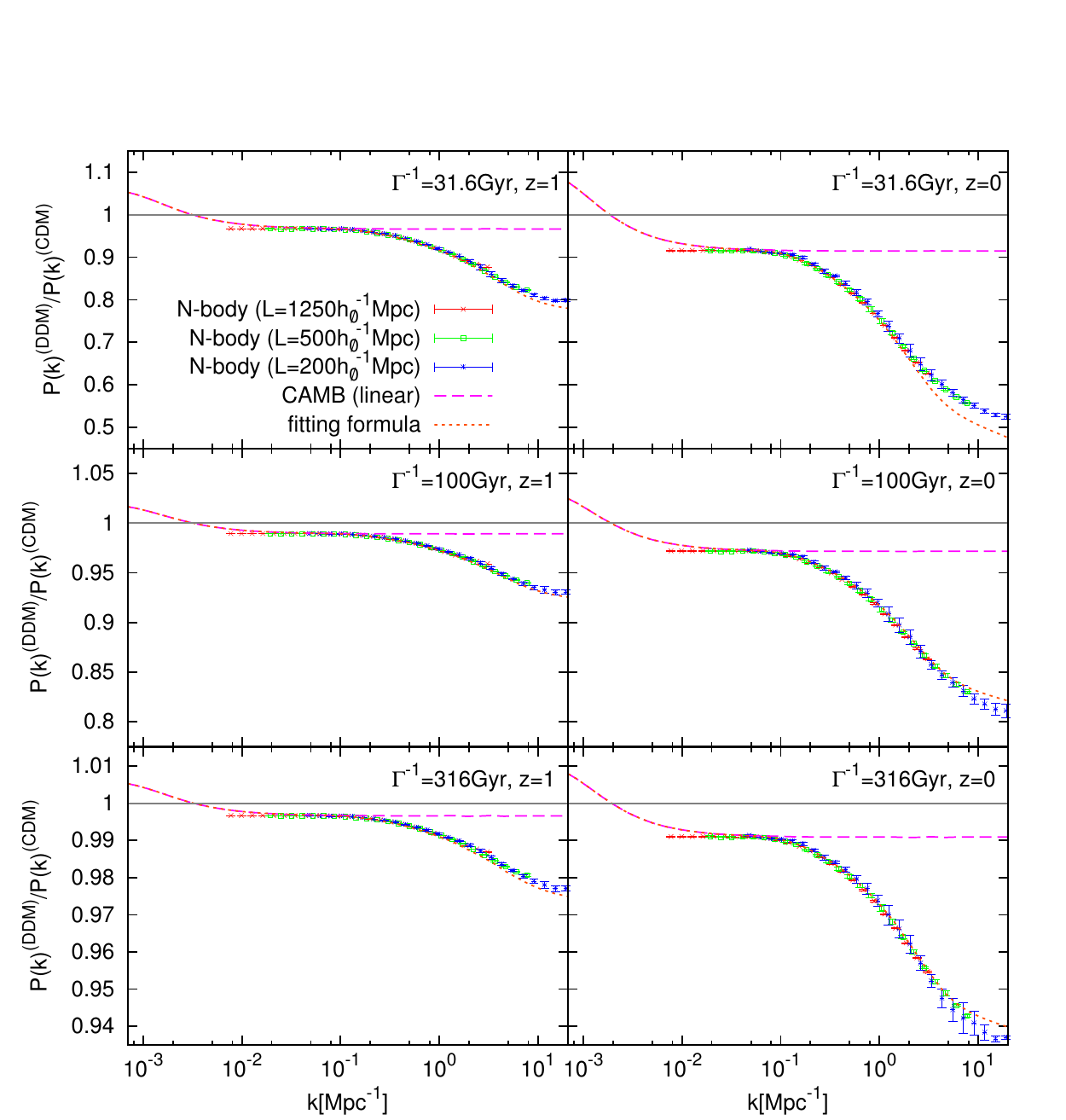}} 
  \end{center}
  \caption{Ratio of non-linear matter power spectrum in the DDM model 
  to that in the CDM model. 
  Red, green and blue points with error bars
  are obtained from N-body simulations with box sizes $L=1250$, $500$ and 200~$h_\emptyset^{-1}$Mpc, respectively.
  The dashed magenta line is obtained from the linear perturbation calculation in the synchronous gauge by using {\tt CAMB}.
  The dotted orange line is obtained by applying our fitting formula in Appendix~\ref{app:fit}, 
  which exactly overlaps with magenta line on large scales.  The solid grey horizontal line indicates unity.
}
  \label{fig:suppression}
\end{figure}

Fig.~\ref{fig:suppression} shows the ratio of the matter power spectrum 
in the DDM model with the lifetimes $\Gamma^{-1}=31.6$, 100 and 316~Gyr to that in the CDM model at redshifts $z=0$ and 1.
The power spectrum of N-body simulations 
is computed using the {\tt ComputePK} code~\cite{2014ascl.soft03015L}\footnote{
http://aramis.obspm.fr/\~{}lhuillier/codes.php
}
with the count-in-cell scheme, and the linear perturbation calculations are performed by using a modified version of 
the {\tt CAMB} code as described in Section~\ref{sec:perturbation}.
Note that here we only take into account the 
overdensity of non-relativistic matter (i.e., baryons and DM), so 
DR is not included in computing the fractional overdensity.
We also note that the units of horizontal axis are Mpc$^{-1}$, which is
different from the ordinary convention of $h\,$Mpc$^{-1}$.
This allows us to extract the difference in growth of density fluctuation, 
after removing the effects of the background expansion.

From Fig.~\ref{fig:suppression}, we can see that the ratio $P^{\rm (DDM)}(k)/P^{\rm (CDM)}(k)$
from the linear calculation and that from the N-body simulations 
coincide in the range $k\simeq0.01-0.1$~Mpc$^{-1}$, where 
the both the linear and Newtonian approximations should be valid. 
Since the effects of perturbations in the massless decay products become less relevant at smaller scales, 
this agreement validates our treatment of the decay products in the N-body simulations.

The ratio $P^{\rm (DDM)}/P^{\rm (CDM)}$ in the linear calculation tends to increase
as $k$ decreases below 10$^{-2}$~Mpc$^{-1}$. This is because the decay product does not free-stream 
until perturbation scales cross the horizon, and its density perturbation more or less 
traces that of other non-relativistic matter (see, e.g., Ref.~\cite{Aoyama:2014tga} for detailed discussion).
This effect is omitted in the N-body simulations, which results in no increase in the ratio on large scales.

As the scales become smaller, the non-linear matter power spectrum shows a larger deviation of the DDM model from the CDM case, while for the linear calculation the ratio stays almost constant.
 This is because the non-linear matter power spectrum is a non-linear function of linear power spectrum
(with mode-mode couplings) and hence small deviations in linear perturbations are in general enhanced 
in the non-linear power spectrum.
As non-linearity becomes more prominent on smaller scales, the deviation becomes more enhanced.

It is also apparent that, except for changes in the overall magnitude,
the fractional deviation $P^{\rm (DDM)}/P^{\rm (CDM)}-1$ has almost the same scale-dependence 
for various values of $\Gamma^{-1}$ and $z$.
This allows us to obtain a fitting formula for the non-linear matter power spectrum in 
the DDM model in Appendix~\ref{app:fit}, which is sufficiently accurate for a wide range of $\Gamma^{-1}$ and $z$. 
This fitting formula is shown by the orange lines in Fig.~\ref{fig:suppression}.
Our formula can reproduce the fractional deviation $P^{\rm (DDM)}/P^{\rm (CDM)}-1$ from 
the N-body results to within an accuracy of 10~\%.

One caveat is that our fitting formula is not tested 
against cosmological parameters other than the fiducial set described above. 
However, as shown in Fig.~\ref{fig:enhancement} in Appendix~\ref{app:fit}, 
the ratio $\epsilon_{\textrm{non-linear}}/\epsilon_{\rm linear}(k)$, where $\epsilon(k) = 1-P^{\rm (DDM)}/P^{\rm (CDM)}$,
does not vary much with redshift or $\Gamma$. This suggests that our fitting formula 
will also depend only weakly on cosmological parameters.
In addition, for the cosmic shear power spectrum, which measures the gravitational potential but not the fractional overdensity of matter,
the impact of uncertainties in the fitting formula would further decrease. 
This is because deviations of the DDM model from the CDM case also arise from the decrease in mean energy density, which suppresses the 
shear power spectra additionally by $\simeq \{(1-r_{dm})+r_{dm} e^{-\Gamma t}\}^2\simeq 1-2r_{dm}\Gamma t$.
The suppression from this effect is in general larger than the contribution from the power spectrum of fractional matter overdensity.
Thus the weak dependence of our fitting formula on cosmological parameters should not significantly affect 
the following analysis.

\section{Constraints from CMB and weak lensing measurements }
\label{sec:constraints}

Having obtained a fitting formula for the non-linear matter power spectrum in the DDM model, 
we now compare the model predictions to the data 
on weak lensing,  CMB and BAO, using a modified {\tt CosmoMC} code~\cite{Lewis:2002ah}.
To obtain the non-linear power spectrum in the DDM model, we combine our fitting formula for 
the $P^{\rm (DDM)}/P^{\rm (CDM)}$ ratio with
the improved {\tt HALOFIT} formula of Ref.~\cite{Takahashi:2012em}. 

For the CMB data, we use the Planck 2013 temperature maps~\cite{Ade:2013kta}
combined with the WMAP 9-year polarization data~\cite{Bennett:2012zja}, and denote the combination as 
CMB. For weak lensing data, we use the tomographic analysis of the CFHTLens data by Ref.~\cite{Heymans:2013fya}, 
which we denote as WL. In particular, we adopt the tomographic cosmic shear power spectrum from blue galaxy samples 
with $N_t=6$  redshift bins and the ``conservative" cut on small angles to mitigate systematic errors from baryonic effects.
We refer the reader to Ref.~\cite{Heymans:2013fya} for more details.
We also use the measurements of baryon acoustic scales from galaxy 
surveys~\cite{Beutler:2011hx,Anderson:2012sa,Padmanabhan:2012hf}
and the Planck CMB lensing power spectrum~\cite{Ade:2013tyw}, which are denoted as BAO and lensing, respectively.
We use four different combinations of these data, which are WL+BAO, 
CMB alone, CMB+WL, and CMB+WL+BAO+lensing.
For the WL+BAO combination, we impose Gaussian priors $\omega_b=0.0223\pm0.0009$
and $n_s=0.96\pm0.02$. 
Here the ranges for the Gaussian priors are taken to be somewhat broader than the
1$\sigma$ errors in the Planck measurements of these quantities. 
However, the prior range does not affect our final results much since WL and BAO are 
not sensitive to $\omega_b$ and $n_s$. 
(In fact, even if we fix the values of $\omega_b$ and $n_s$ in obtaining the 
constraints from WL+BAO, the results do not change much.)

The cosmic shear power spectrum is obtained as follows.
Using the Limber approximation, 
the cross-angular power spectrum of the convergence fields $\kappa$ at redshifts $z_1$ and $z_2$ 
at multipole $\ell$ is given as (see, e.g., Ref.~\cite{Bartelmann:1999yn})
\begin{equation}
C^\kappa_\ell(z_1, z_2)=\int dr 
\left(1-\frac{r}{r_1}\right)
\left(1-\frac{r}{r_2}\right)
\left[4\pi Ga^2\bar \rho(a)\right]^2
P(k=\frac{\ell}{r}, z), 
\label{eq:conv}
\end{equation}
where $r(z)$ is the comoving distance to redshift $z=1/a-1$, and $r_i=r(z_i)$, $i=1,2$. 
Note that we have omitted the contribution from perturbations of the decay products, because
current observations are sensitive to perturbations at sub-horizon scales, where
those of decay products can be neglected.
By applying the fitting formula for the non-linear matter power spectrum in the DDM model, 
Eq.~\eqref{eq:conv} can provide the theoretical prediction for weak lensing observations.

In order to avoid confusion, let us remind the reader of the definitions of parameters
which we constrain in the following. 
$\omega_{dm}$ is the initial density parameter
of dark matter and in general differs from one at present, $\Omega_{dm}h^2$;
$\sigma_8$ is computed from the matter power 
spectrum of non-relativistic matter; $H_0=100h$ is the actual Hubble parameter 
at present and should not be confused with $100h_\emptyset$. 
The definitions of these parameters are presented in Sec. \ref{sec:background}.

\begin{figure}
  \begin{center}
      \hspace{0mm}\scalebox{1.}{\includegraphics{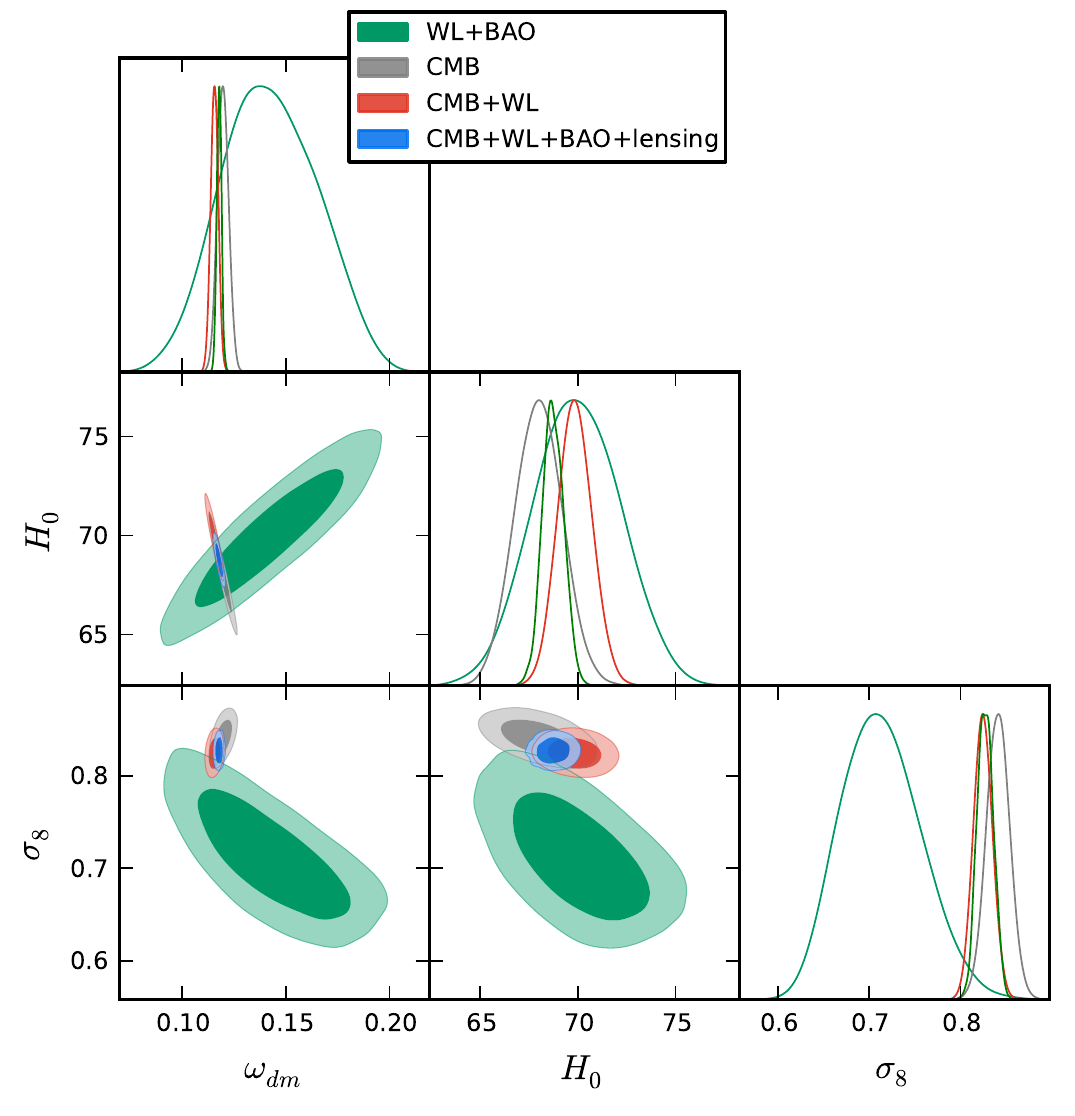}} 
  \end{center}
  \caption{
  Constraints on the CDM model. The 1d and 2d posterior distributions for parameters 
  $\omega_{dm}$, $H_0$ and $\sigma_8$ are shown, with the 2d constraints given at 68~\% and 95~\% C.L.
  Green, grey, red and blue lines correspond to the data sets 
  WL, CMB, CMB+WL and CMB+WL+BAO+lensing, respectively.
  }
  \label{fig:CDM}
\end{figure}

Let us take a look at the parameter constraints in the CDM model first. Fig.~\ref{fig:CDM} shows 
the constraints on parameters of particular interest, $\omega_{dm}$, $\sigma_8$ and $H_0$.
In the $\omega_{dm}$-$\sigma_8$ and $H_0$-$\sigma_8$ planes, 
the posterior distributions from the CMB and WL+BAO datasets 
overlap only marginally at the 95~\% C.L. level.
In particular, the tension resides in estimation of $\sigma_8$.
This result highlights the existence of the tension between Planck and CFHTLens, 
which has been argued in the literature~\cite{Ade:2013zuv,MacCrann:2014wfa,Planck:2015xua,Petri:2015ura}.
To be more quantitative, we explore 
the posterior distributions from these two datasets in the 3-dimensional parameter space of 
($\omega_{dm}$, $\sigma_8$, $H_0$). We find that 
the posterior distributions overlap only at more than 90~\% C.L.,
which is broadly consistent with the results of Ref.~\cite{MacCrann:2014wfa},
where the full 6-dimensional parameter space is analysed.

\begin{figure}
  \begin{center}
      \hspace{0mm}\scalebox{1.}{\includegraphics{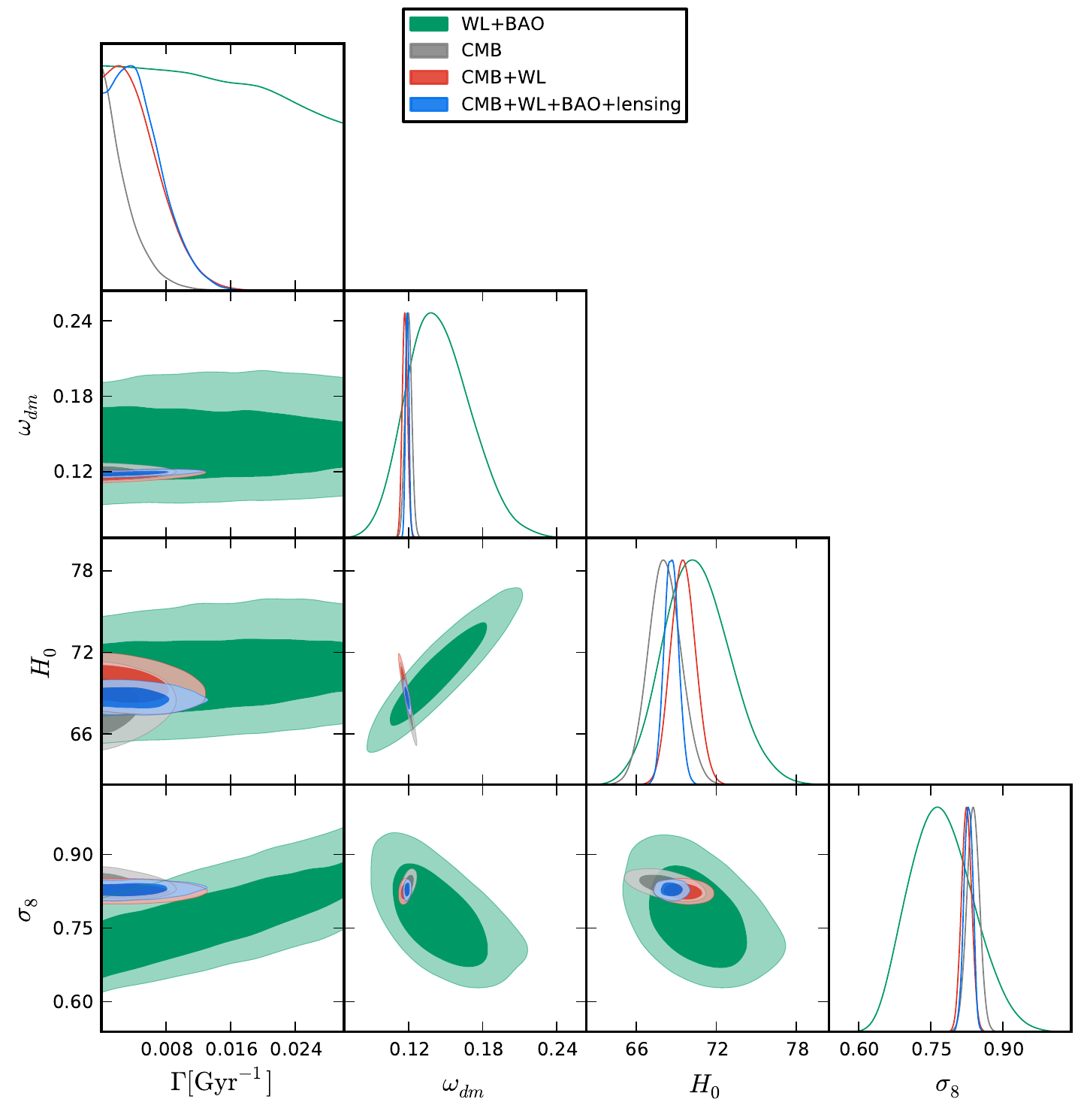}} 
  \end{center}
  \caption{
  The same as in Fig.~\ref{fig:CDM}, but for the DDM model, 
  which has an additional parameter $\Gamma$.
  }
  \label{fig:DDM}
\end{figure}

Fig.~\ref{fig:DDM} shows parameter constraints in the DDM model, which has an extra parameter $\Gamma$.
Compared to the CDM case, the WL+BAO dataset allows a larger value of $\sigma_8$.
This leads to a substantial reduction of the 
tension between Planck and CFHTLens in the CDM model. 
In the same manner as in the CDM model, we examine the overlap of the posterior distributions from 
CMB and WL+BAO in the 4-dimemsional parameter space ($\Gamma$, $\omega_{dm}$, $\sigma_8$, $H_0$) of the DDM model, 
and find that they overlap at 55~\% C.L. 
This level of improvement appears somewhat better than that achieved by the addition of massive (sterile)
neutrinos~\cite{MacCrann:2014wfa}, 
which reports that posterior distributions from CFHTLens and CMB overlap at 90~\% C.L.
in the massive neutrino model, that has one extra parameter, 
while they do so at 64~\% C.L. in the sterile neutrino model, that has two extra parameters. 
However, a direct comparison is complicated by the fact that the analysis
uses the shear power spectrum data on smaller angular scales than we do, 
while additionally accounting for theoretical uncertainties due to intrinsic alignment of 
galaxies and the effects of AGN feedback. 

We would like to comment on the degeneracy between $\sigma_8$ and $\Gamma$ from the 
WL+BAO dataset. 
As seen from the bottom-left panel in Fig.~\ref{fig:DDM}, these two parameters are positively 
correlated, which may at first look counterintuitive as matter fluctuations should be
more suppressed as $\Gamma$ becomes larger. However, $\sigma_8$
represents the amplitude of the overdensity in non-relativistic matter (baryon+DM), which is less suppressed 
than that in the total matter (i.e. baryon+DM+DR) or gravitational potential.
Therefore, as $\Gamma$ is increased, $\sigma_8$ should be also increased
to fit the amplitude of the cosmic shear power spectrum.

Finally, we present constraints on the lifetime of DDM, $\Gamma^{-1}$, from the various datasets. 
From the CMB alone, we obtain $\Gamma^{-1}\ge140$ Gyr at 95~\% C.L. This is very similar to the constraints in 
Ref.~\cite{Audren:2014bca}, where the authors use CMB+BAO combined with the galaxy power spectrum.
Combining CMB with WL, we obtain a less stringent constraint, $\Gamma^{-1}\ge97$ Gyr.
This is because the CMB+WL data prefers a nonzero value of $\Gamma$, in order to mitigate 
the $\sigma_8$ tension in the CDM model. Inclusion of BAO and the CMB lensing data changes
the lower bound only by a few percent. 
On the other hand, WL data alone is not yet sufficiently constraining even when it is
combined with BAO, so we cannot obtain meaningful constraints on $\Gamma^{-1}$ above $\Gamma^{-1}\ge31.6$ Gyr, 
where we can guarantee the accuracy of our fitting formula.


\section{Conclusion}\label{sec:conclusion}

We have studied cosmological structure formation in a model of dark matter
decaying into massless invisible particles (``dark radiation"), denoted as DDM model. Employing N-body simulations, 
we have obtained an accurate fitting formula for the non-linear matter power spectrum
in the DDM model. Using this fitting formula, we compared predictions of
DDM to the tomographic cosmic shear power spectrum measurements from the CFHTLens survey
in conjunction with CMB and BAO data. We have demonstrated that the tension present
in estimations of $\sigma_8$ in the CDM model is substantially 
alleviated in the DDM model.  

Quantitatively, we find that the posterior distributions
from two different datasets, CMB alone and WL+BAO, overlap with each other
at 55~\% C.L. in the DDM model, while they do so only at 90~\% C.L. in the CDM model.
We thus suggest the $\sigma_8$ discrepancy may be hinting at the possibility that dark matter could be unstable.
On the other hand, current cosmological observations do not provide powerful enough constraints to 
nail down the finite lifetime of DM, $\Gamma^{-1}$. Combining the CFHTLens weak lensing data 
with the Planck and WMAP CMB data, we obtain a lower bound, $\Gamma\ge 97$~Gyr.
This is less stringent than the one from CMB alone, $\Gamma\ge 140$~Gyr, which reflects the 
preference for a non-zero dark matter decay width from CMB+WL.

In the near future weak lensing measurements such as  the Subaru Hyper Suprime-Cam~\cite{HSC},
the Dark Energy Survey~\cite{Abbott:2005bi}, and Euclid~\cite{Laureijs:2011gra}, should increase the statistics of the data
by orders of magnitude. The DDM scenario we have considered in this paper will thus definitely be tested
by these future observations. One should also point out that certain cosmological observations other than weak lensing 
 may also show signs of tension with the CMB and the $\Lambda$CDM model.
In particular, the number of galaxy clusters observed by the Sunyaev-Zeldovich effect and X-ray observations
are known to be less than the prediction of the best-fit $\Lambda$CDM model to the CMB power spectrum~\cite{Ade:2013zuv}. 
Since the amplitude of matter fluctuations are suppressed in the DDM model, the abundance of dark matter haloes and hence hosted galaxy clusters
also becomes less than in the CDM model.
It would be interesting to consider whether the DDM model can in a consistent manner
resolve the possible disagreements between all such observations. We leave this discussion to a future work.

\acknowledgments 

T.S. would like to thank Naoshi Sugiyama for valuable comments and suggestions.
T.T. would like to thank the Helsinki Institute of Physics for the hospitality
during the visit, where a part of this work has been done.
The work of T.T. is partially supported by the Grant-in-Aid for Scientific
research from the Ministry of Education, Science, Sports, and
Culture, Japan, No.~23740195. We thank the CSC - IT Center 
for Science (Finland) for computational resources.

\appendix

\section{Approximations in solving Boltzmann hierarchy}
\label{app:approx}

In this appendix, we summarise the approximations used 
in solving the Boltzmann hierarchy of Eq.~\eqref{eq:brightness} 
in the linear perturbation calculation. Our approximations are 
based on Ref.~\cite{Blas:2011rf}.

First of all, let us define a new variable $\tilde X\equiv \tilde\omega_{dr} X$.
Then Eq.~\eqref{eq:brightness0} can be rewritten as
\begin{eqnarray}
\dot{\tilde X}+ik\mu \tilde X=
\tilde\omega_{dr}\left\{\dot\eta_T-\frac12(\dot h_L+6\dot \eta_T) \mu^2\right\}
+\dot{\tilde\omega}_{dr}\frac{\delta_{dm}}4.
\label{eq:tildeX}
\end{eqnarray}
Noting that $\tilde X_l$ vanishes at $\tau=0$, 
the formal solution of this equation can be given as 
\begin{equation}
\tilde X(\tau)=\int^\tau_0 d\tau' e^{-ik\mu(\tau-\tau')}
\left\{A(\tau')+B(\tau')\mu^2\right\}, 
\end{equation}
where 
\begin{eqnarray}
A&=&\tilde\omega_{dr}\dot\eta_T+\dot{\tilde\omega}_{dr}\frac{\delta_{dm}}4, \\
B&=&-\frac{\tilde\omega_{dr}}2\left(\dot h_L+6\dot\eta_T\right).
\end{eqnarray}
Since $B$ vanishes at $\tau=0$, 
after integrating by part and multipole expansion, we obtain
\begin{eqnarray}
&&\tilde X_l(\tau)-\frac1{k^2}\left(\dot B(\tau)\delta_{l0}+\frac k3B(\tau)\delta_{l1}\right) \notag\\
&&\quad=-\frac{\dot B(0)}{k^2}j_l(k\tau) 
+\int^\tau_0 d\tau' j_l(k(\tau-\tau'))\left(A(\tau')-\frac{\ddot B(\tau')}{k^2}\right), 
\label{eq:formal}
\end{eqnarray}
where $j_l$ is the first kind of spherical Bessel function.

After horizon crossing $k\tau\simeq1$, $\tilde X_l$ starts to oscillate as the free-streaming 
generates higher multipole moments from lower ones. In this regime, the {\it ultra-relativistic fluid approximation} (UFA)
developed in Ref.~\cite{Blas:2011rf} can be applied.
Following Ref.~\cite{Blas:2011rf}, let us introduce a new variable 
$Y_l=\tilde X_l-\frac1{k^2}(\dot B\delta_{l0}+\frac k3B\delta_{l1})$, whose behaviour is quite close to that of $j_l(k\tau)$, 
which satisfies $j'_l(x)-j_{l-1}(x)+\frac{l+1}xj_l(x)=0$.
From Eq.~\eqref{eq:formal}, one can derive the following equation:
\begin{eqnarray}
&&\dot Y_l-kY_l+\frac{l+1}\tau Y_l \notag \\ 
&&\quad=\left\{A(\tau)-\frac{\ddot B}{k^2}(\tau)\right\}\delta_{l0}
-\frac{l+1}\tau \int^\tau_0d\tau' \frac{\tau'}{\tau-\tau'}j_l(k(\tau-\tau'))\left\{A(\tau')-\frac{\ddot B(\tau')}{k^2}\right\}.
\label{eq:UFA}
\end{eqnarray}
The integrand on the r.h.s becomes large only at
$k\tau-l \lesssim k\tau'\lesssim k\tau$. This allows one to approximate the third term on the r.h.s by
\begin{equation}
-(l+1)\left\{A(\tau)-\frac{\ddot B(\tau)}{k^2}\right\}\int^\infty_0 dx \frac{j_l(x)}x.
\end{equation}
In particular, for $l=2$, Eq.~\eqref{eq:UFA} can be approximately rewritten in terms of $X_l$ as
\begin{equation}
\dot X_2=
kX_1-\left(\frac3\tau+\frac{\dot{\tilde\omega}_{dr}}{\tilde\omega_{dr}}\right)X_2
+\frac{\dot h_L}6+2\dot\eta_T+\frac{\dot{\tilde\omega}_{dr}}{\tilde\omega_{dr}}\frac{\delta_{dm}}4,
\label{eq:ell2}
\end{equation}
where we have assumed $|\ddot B/k^2|\sim |B/(k\tau)^2|\ll |A|$.

On the other hand, deeper inside horizon, perturbations in $X_l$ become very small
and their effects on the perturbation evolution of non-relativistic matter are not relevant unless
the energy density of the decay product dominates the Universe. On such scales, 
the oscillation period of $X_l$ is much shorter than the typical time-scale of perturbation evolution
of non-relativistic matter, so that only non-oscillatory components of $X_l$ suffice in computing the matter perturbation evolution.
In this regime, the {\it radiation streaming approximation} (RSA) in Ref.~\cite{Blas:2011rf} can be applied.
Let us consider the Boltzmann hierarchy for $\tilde X_l$: 
\begin{eqnarray}
\dot{\tilde X}_0&=&-k\tilde X_1-\tilde \omega_{dr}\frac{\dot h_L}6+\dot{\tilde\omega}_{dr}\frac{\delta_{dm}}4, 
\label{eq:RSA0}
\\
\dot{\tilde X}_1&=& \frac{k}3\tilde X_0,
\label{eq:RSA1}
\end{eqnarray}
where we have assumed $\tilde X_l=0$ for $l\ge2$. Combining 
these two equations we obtain
\begin{equation}
X_0=\frac1{\tilde\omega_{dr} k^2}\frac\partial{\partial\tau}
\left(
-\tilde\omega_{dr}\frac{\dot h_L}2
+\dot{\tilde\omega}_{dr}\frac{3\delta_{dm}}4
\right),
\label{eq:RSA_X0}
\end{equation}
where we have assumed $|\ddot{\tilde X}_0|\ll k^2|\tilde X_0|$ since we are interested in
the non-oscillatory components of $\tilde X_l$. By integrating Eq.~\eqref{eq:RSA0}, we then obtain
\begin{equation}
X_1=-\frac{\dot h_L}{6k}+\frac{\dot{\tilde\omega}_{dr}}{\tilde\omega_{dr}}\frac{\delta_{dm}}{4k},
\label{eq:RSA_X1}
\end{equation}
where we have used the fact that $\tilde X_l$, $\tilde\omega_{dr}\dot h_L$ and $\dot{\tilde\omega}_{dr} \delta_{dm}$ vanish at $\tau=0$.
To compute $\ddot h_L$, we adopt the same procedure as in Ref.~\cite{Blas:2011rf}, which is valid in a matter-dominated Universe.

In our calculation, the Boltzmann hierarchy Eq.~\eqref{eq:brightness} 
truncated at $l=l_{\rm max}$ is directly solved from superhorizon scales up to 
$k\tau=l_{\rm max}$. Then we switch the UFA on and the Boltzmann hierarchy is truncated at $l=2$
by adopting Eq.~\eqref{eq:ell2}. Finally, when $k\tau\ge10\times l_{\rm max}$ in the matter dominated Universe, 
we switch the RSA on and substitute Eqs.~\eqref{eq:RSA_X0}-\eqref{eq:RSA_X1}
in the source term of the Einstein equation.
These approximations avoid the unwanted blowup of $X_l$ caused by the truncation of Boltzmann hierarchy.
Note that to employ the RSA, we need to assume that the decay product is a minor component of the energy density of the Universe.
This assumption is valid in the parameter region we explore in this paper.

\section{Fitting formula for non-linear matter power spectrum in DDM model}
\label{app:fit}
In this appendix we present the fitting formula for the non-linear power spectrum of 
non-relativistic matter in the DDM model. 
First, let us present the fitting formula for the suppression in the linear power spectrum 
$P_{\rm linear}$ at the small scale limit.
We find that $\epsilon_{\rm linear}=1-P^{\rm (DDM)}_{\rm linear}/P^{\rm (CDM)}_{\rm linear}$ at $k\to\infty$ 
can be approximately given as
\begin{equation}
\epsilon_{\rm linear}(\Gamma,z)=
\alpha\left(\frac\Gamma{\mathrm{Gyr}^{-1}}\right)^\beta
\left(\frac1{z+1}\right)^\gamma,
\end{equation}
where $\alpha$, $\beta$ and $\gamma$ are functions of $\omega_b$, $h_\emptyset$ and 
$\omega_m=\omega_b+\omega_{dm}$ in the case of our model, which can be fitted by
\begin{eqnarray}
\alpha(\omega_b,h_\emptyset,\omega_m)
&=&
(5.323-1.4644u-1.391v)
+(-2.055+1.329u+0.8672v)w \notag \\
&&
\quad+(0.2682-0.3509u)w^2, \\
\beta(\omega_b,h_\emptyset,\omega_m)
&=&
0.9260+(0.05735-0.02690v)w
+(-0.01373+0.006713v)w^2, \\
\gamma(\omega_b,h_\emptyset,\omega_m)
&=&
(1.653-0.7860v)
+(0.4884+0.1754v)w\notag \\
&&\quad+(-0.2512+0.07558v)w^2,
\end{eqnarray}
where $u=\omega_b/0.02216$, $v=h_\emptyset/0.6777$ and $w=\omega_m/0.1412$.
The accuracy is 10~\% for $\Gamma^{-1}\ge31$Gyr and $z\le1$
as long as $0.019\le\omega_b\le0.026$, $0.6\le h_\emptyset\le0.8$ and 
$0.09\le\omega_m\le 0.28$. 

\begin{figure}
  \begin{center}
      \hspace{-5mm}\scalebox{1.4}{\includegraphics{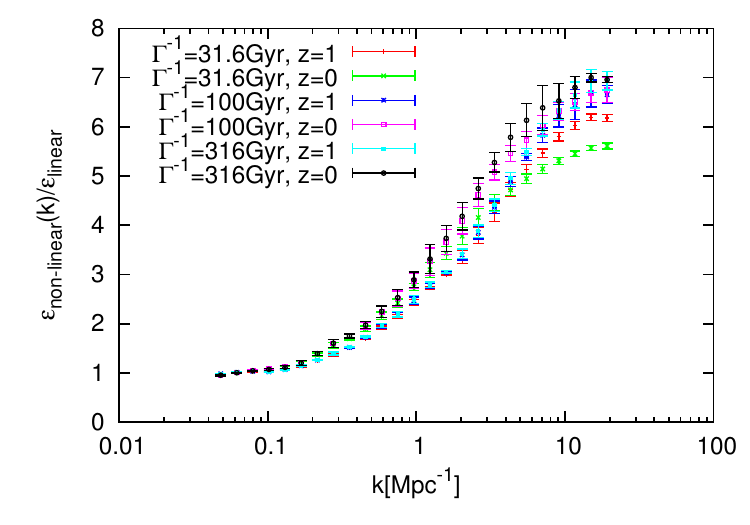}}
  \end{center}
  \caption{
  Enhancement of suppression from linear to non-linear matter power spectrum 
  $\epsilon_{\textrm{non-linear}}(k)/\epsilon_{\rm linear}$ for
  $\Gamma^{-1}=$31.6, 100, 316~Gyr and $z=$0, 1. 
  Here we adopt the result of N-body simulations with a box size $L=200~h_\emptyset^{-1}$Mpc.
  }
  \label{fig:enhancement}
\end{figure}

In Fig.~\ref{fig:enhancement} we show the suppression in the non-linear power spectrum 
$\epsilon_{\textrm{non-linear}}(k)=1-P^{\rm (CDM)}_{\textrm{non-linear}}/P^{\rm (DDM)}_{\textrm{non-linear}}$ 
in units of $\epsilon_{\rm linear}$ for $\Gamma^{-1}=31$, 100, 316~Gyr and $z=$0,~1.
The cosmological parameters in Eq.~\eqref{eq:fiducial} are adopted here.
One can see that $\epsilon_{\textrm{non-linear}}(k)/\epsilon_{\rm linear}$
has only a weak dependence both on $\Gamma^{-1}$ and $z$.
We find $\epsilon_{\textrm{non-linear}}(k)/\epsilon_{\rm linear}$ can be fitted by the following functional form: 
\begin{equation}
\frac{\epsilon_{\textrm{non-linear}}(k)}{\epsilon_{\rm linear}}=
\frac{1+a(k/{\rm Mpc}^{-1})^{p}}
{1+b(k/{\rm Mpc}^{-1})^{q}},
\end{equation}
where $a,~b,~p$ and $q$ are functions of $\Gamma$ and $z$ as 
\begin{eqnarray}
a(\Gamma,z)&=& 0.7208
+2.027\left(\frac\Gamma{{\rm Gyr}^{-1}}\right)
+3.431\left(\frac1{1+z}\right),\\
b(\Gamma,z)&=& 0.0120
+2.786\left(\frac\Gamma{{\rm Gyr}^{-1}}\right)
+0.6499\left(\frac1{1+z}\right),\\
p(\Gamma,z)&=& 1.045
+1.225\left(\frac\Gamma{{\rm Gyr}^{-1}}\right)
+0.2207\left(\frac1{1+z}\right),\\
q(\Gamma,z)&=& 0.9922
+1.735\left(\frac\Gamma{{\rm Gyr}^{-1}}\right)
+0.2154\left(\frac1{1+z}\right).
\end{eqnarray}
The accuracy of this fitting formula is better than 10~\% for $\Gamma^{-1}\ge31$~Gyr and $z\le1$.
We note that this fitting formula is calibrated with simulations only for the fiducial cosmological parameters 
in Eq.~\eqref{eq:fiducial}. However, having observed the fact that $\epsilon_{\textrm{non-linear}}(k)/\epsilon_{\rm linear}$
varies only weakly for various $\Gamma^{-1}$ and $z$, we speculate that
this quantity does not vary much for different cosmological parameters, as long as the
linear power spectrum does not deviate significantly from the fiducial one.
This condition is effectively satisfied when one compares model predictions with observed data
of matter power spectrum or CMB.

\end{document}